\documentclass{aa}
\usepackage{graphicx}
\usepackage{txfonts}
\usepackage{ulem}

\def\Mjup{\hbox{$M_{\rm Jup}$}}
\def\Rjup{\hbox{$R_{\rm Jup}$}}
\def\aRs{\hbox{$a/R_\star$}}
\def\RpRs{\hbox{$R_{\rm p}/R_\star$}}
\def\Tmid{\hbox{$T_{\rm mid}$}}
\def\Teq{\hbox{$T_{\rm eq}$}}

\usepackage[colorlinks=true,citecolor=blue]{hyperref}
\usepackage{natbib}
\begin{document}

   \title{The GTC exoplanet transit spectroscopy survey IX. }

   \subtitle{Detection of haze, Na, K, and Li in the super-Neptune WASP-127b}

   \author{G. Chen\inst{1,2,3}
          \and
          E. Pall\'{e}\inst{1,2}
          \and
          L. Welbanks\inst{4}
          \and
          J. Prieto-Arranz\inst{1,2}
          \and
          N. Madhusudhan\inst{4}
          \and
          S. Gandhi\inst{4}
          \and
          N. Casasayas-Barris\inst{1,2}
          \and
          F. Murgas\inst{1,2}
          \and
          L. Nortmann\inst{1,2}
          \and
          N. Crouzet\inst{1,2}
          \and
          H. Parviainen\inst{1,2}
          \and
          D. Gandolfi\inst{5}
          }

   \institute{Instituto de Astrof\'{i}sica de Canarias, V\'{i}a L\'{a}ctea s/n, E-38205 La Laguna, Tenerife, Spain\\
         \email{gchen@iac.es}
         \and
             Departamento de Astrof\'{i}sica, Universidad de La Laguna, Spain
         \and
             Key Laboratory of Planetary Sciences, Purple Mountain Observatory, Chinese Academy of Sciences, Nanjing 210008, China
         \and
             Institute of Astronomy, University of Cambridge, Madingley Road, Cambridge CB3 0HA, UK
         \and
             Dipartimento di Fisica, Universit\'{a} di Torino, Via P. Giuria 1, I-10125, Torino, Italy
             }

   \date{Received March 16, 2018; accepted May 18, 2018}

 
  \abstract
  {Exoplanets with relatively clear atmospheres are prime targets for detailed studies of chemical compositions and abundances in their atmospheres. Alkali metals have long been suggested to exhibit broad wings due to pressure broadening, but most of the alkali detections only show very narrow absorption cores, probably because of the presence of clouds. We report the strong detection of the pressure-broadened spectral profiles of Na, K, and Li absorption in the atmosphere of the super-Neptune WASP-127b, at 4.1$\sigma$, 5.0$\sigma$, and 3.4$\sigma$, respectively. We performed a spectral retrieval modeling on the high-quality optical transmission spectrum newly acquired with the 10.4~m Gran Telescopio Canarias (GTC), in combination with the re-analyzed optical transmission spectrum obtained with the 2.5~m Nordic Optical Telescope (NOT). By assuming a patchy cloudy model, we retrieved the abundances of Na, K, and Li, which are super-solar at 3.7$\sigma$ for K and 5.1$\sigma$ for Li (and only 1.8$\sigma$ for Na). We constrained the presence of haze coverage to be around 52\%. We also found a hint of water absorption, but cannot constrain it with the global retrieval owing to larger uncertainties in the probed wavelengths. WASP-127b will be extremely valuable for atmospheric characterization in the era of James Webb Space Telescope.}


   \maketitle
%

\section{Introduction}\label{sec:intro}

The characterization of exoplanet atmospheres could play a critical role in connecting to planet formation histories \citep[e.g.,][]{2011ApJ...743L..16O,2014ApJ...794L..12M,2016ApJ...832...41M}. The inference of atmospheric metallicity and elemental ratios must rely on the diagnosis of spectral features originated in planetary atmospheres. Transmission spectroscopy has taught us that clouds and hazes are common in hot Jupiter atmospheres \citep[e.g.,][]{2016Natur.529...59S}, which mute spectral features and lead to degeneration in the parameter space. Nevertheless, great effort has been made with the WFC3 instrument on the Hubble Space Telescope (HST) to search for the water feature within 1.1--1.7~$\mu$m. Based on the water abundance, an atmospheric metallicity enrichment trend has been noticed from massive hot Jupiters to warm Neptunes \citep[e.g.,][]{2014ApJ...793L..27K,2017Sci...356..628W,2018ApJ...855L..30A,Nikolov2018bNaturePaperWASP96b}.  

\object{WASP-127b} is one of the rare short-period super-Neptunes in the transition gap from Jupiter mass to Neptune mass \citep{2016A&A...589A..75M}, and the characterization of its atmosphere could help explain its formation mechanisms. This object has a mass of $0.18\pm0.02$~\Mjup\ and a radius of $1.37\pm0.04$~\Rjup, and orbits a G5 star every 4.17 days \citep{2017A&A...599A...3L}. Its large atmospheric scale height $H_\mathrm{eq}/R_\star=kT/(\mu gR_\star)=0.00243$ (assuming $\Teq =1400$~K, $\mu = 2.3$~g~mol$^{-1}$, $g_\mathrm{p}=2.14$~m~s$^{-1}$, $R_\star=1.39$~$R_\sun$), together with its bright host star ($V=10.2$), makes it one of the most ideal targets for atmospheric characterization. \citet{2017A&A...602L..15P} studied its atmosphere via transmission spectroscopy with the ALFOSC spectrograph at the 2.5~m Nordic Optical Telescope (NOT), and found evidence of a Rayleigh scattering slope at the blue optical, a hint of Na absorption, and suspicious signal attributed to TiO/VO absorption. 

We report the detection of scattering haze, Na, and K, and a hint of water based on the new data acquired with the 10.4~m Gran Telescopio Canarias (GTC) together with the published NOT data. This paper is organized as follows. In Sect.~\ref{sec:data}, we summarize the observations and data reduction. In Sect.~\ref{sec:analysis}, we detail the light-curve analysis. In Sect.~\ref{sec:discuss}, we discuss the atmospheric properties inferred from the transmission spectrum. We present additional figures and tables in the Appendices. 

\section{Observations and data reduction}
\label{sec:data}

\begin{figure}[h!]
\centering
\includegraphics[width=1\linewidth]{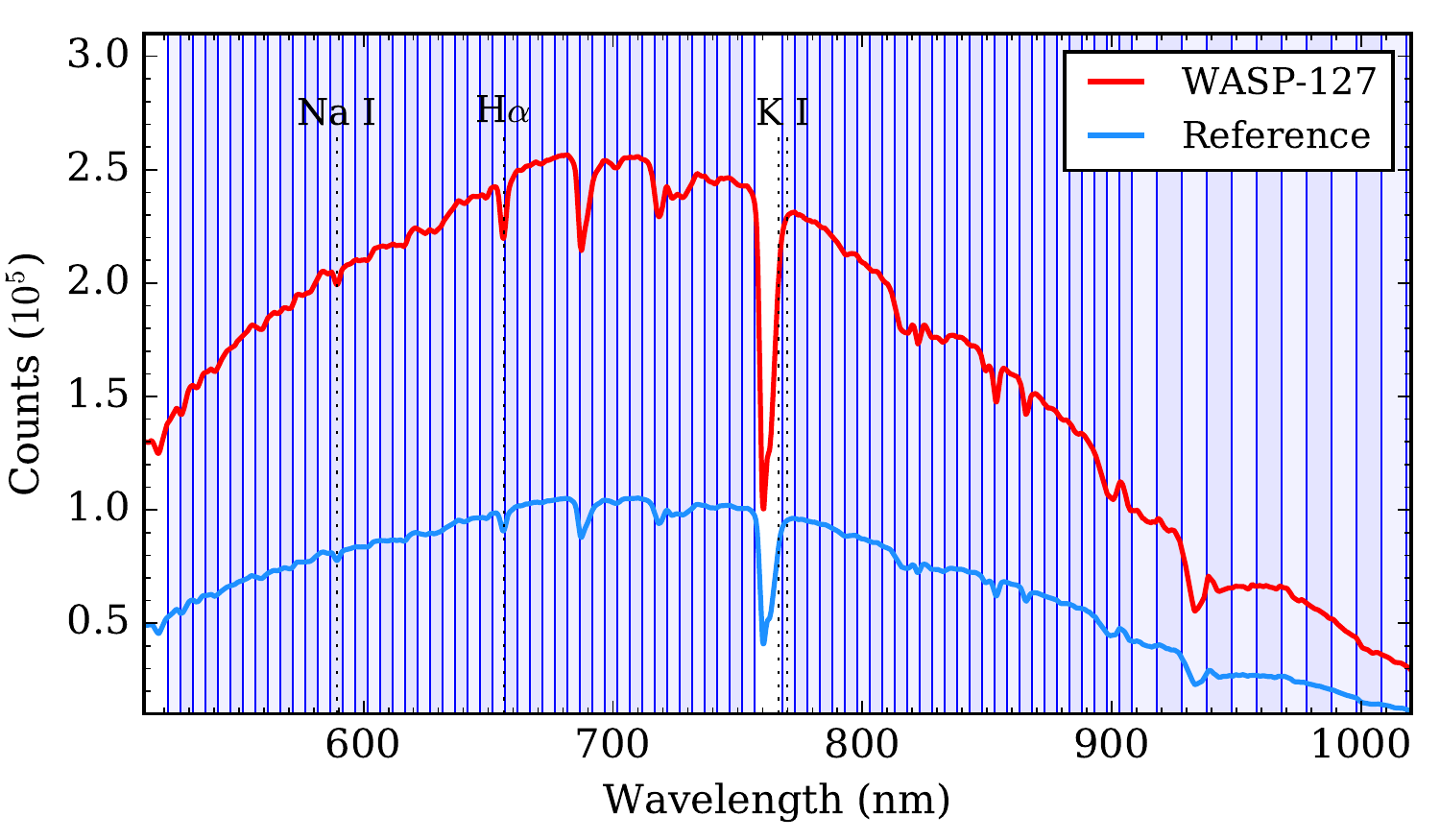}
\caption{Example stellar spectra of \object{WASP-127} (red) and the reference star (blue) obtained with the R1000R grism of GTC/OSIRIS on the night of January 19, 2018. The color-shaded areas indicate the divided passbands that are used to create the spectroscopic light curves.\label{fig:GTCSpectra}}
\end{figure}

One transit of \object{WASP-127b} was observed on the night of January 19, 2018 with the OSIRIS spectrograph \citep{2012SPIE.8446E..4TS} at the GTC. The CCD chip 1 of OSIRIS was configured with $2\times2$ binning (0.254$''$ per pixel) and 200~kHz readout mode to record the spectral data, while chip 2 was switched off. The spectral data were collected using the R1000R grism and a customized 40$''$-wide slit. A reference star (\object{TYC 4916-897-1}; $r'\mathrm{mag}=11.0$) at a separation of 40.5$''$ was simultaneously monitored with \object{WASP-127} ($r'\mathrm{mag}=10.0$).

The observation lasted from 00:26 UT to 07:16 UT, while the morning twilight started at 06:42 UT. Two jumps in star locations occurred because of guiding problems, and the stars were roughly put back to the original location afterward (see Appendix \ref{app:guiding}). Three exposure times were tested in the first 36 exposures, and fixed to 6 sec for the remaining 777 exposures until the end. The duty cycle is roughly 20\% owing to the readout overheads of $\sim$23 sec. The night was dark and mostly clear. The airmass decreased from 2.01 to 1.19 and then rose to 1.89 in the end. The observation was slightly defocused and the spatial point spread function is well Gaussian. The measured full width at half maximum (FWHM) of the spatial profile varied between 1.25$''$ and 3.12$''$, resulting in a seeing-limited spectral resolution of $\sim$20~\AA.

The two-dimensional spectral images were calibrated for overscan, bias, flat field, and sky background following the method described in \citet{2017A&A...600A.138C,2017A&A...600L..11C}. The sky background model was constructed in the wavelength space, where no curvature of sky lines exists, and was then subtracted from the original spectral image after being transformed back to the original pixel space. This process made use of wavelength solutions that were created using the line lists from the HeAr, Ne, and Xe arc lamps, which were acquired with the R1000R grism and the 1.3$''$ slit. The one-dimensional spectra were extracted using the optimal extraction algorithm \citep{1986PASP...98..609H} that has a fixed aperture diameter of 21 pixels ($\sim$5.3$''$). This diameter value has been optimized over a wide range of aperture sizes and results in the least scatter in the white-color light curve. 

To create light curves, the UT time stamp was shifted to each mid-exposure time and converted to the Barycentric Julian Date in the Barycentric Dynamical Time standard \citep[BJD$_\mathrm{TDB}$;][]{2010PASP..122..935E}. The flux of each star at each exposure was integrated over any given wavelength range, where the counts of the two edge pixels were fractionally added and those of the in-between pixels were directly summed. The light curve was recorded as the flux ratios between the target and reference stars after being normalized by the out-of-transit data points. The white-color light curve was integrated over 535--908~nm, but excluding the 755--768~nm region to minimize the contamination of the strong telluric O$_2$-A band. The spectroscopic light curves were created to have a wavelength span of 5~nm at wavelengths shorter than 908~nm and 10~nm at longer wavelengths (see Fig.~\ref{fig:GTCSpectra}).

To avoid ruining the systematics training process, the data taken during the big jump, when the first guiding loss occurred, were not used in the subsequent analyses. A few exposures at the beginning and end of the observations were not used because they are either at very high airmass ($\sim$2) or in the late phase of morning twilight (see the gray shaded areas in Fig.~\ref{fig:GTCTrends}).

\section{Light-curve analysis}\label{sec:analysis}

\begin{figure}
\centering
\includegraphics[width=1\linewidth]{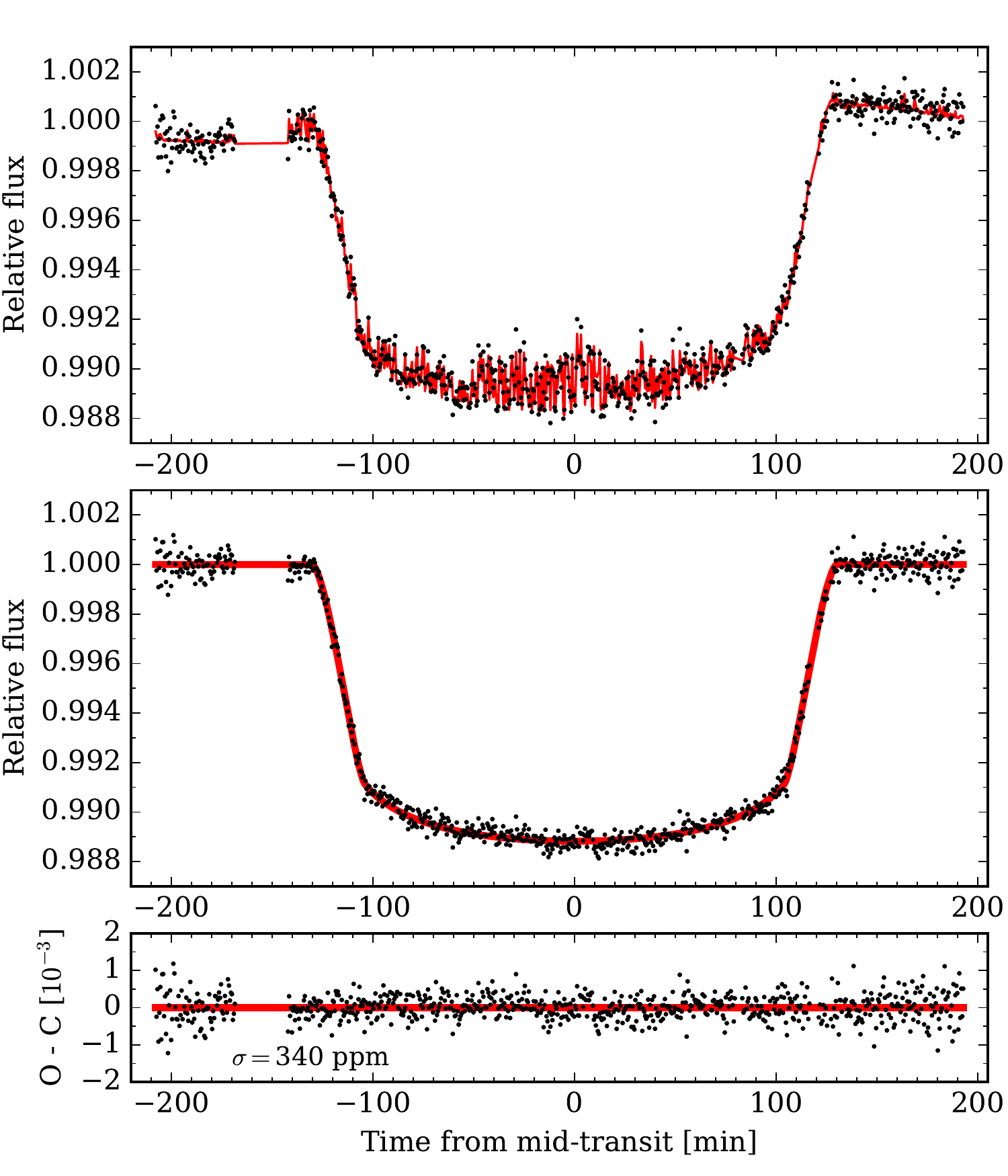}
\caption{White-color light curve obtained with GTC/OSIRIS. From top to bottom are the raw light curve (i.e., the normalized target-to-reference flux ratios), the light curve corrected for systematics, and the best-fitting residuals. The red line shows the best-fitting model.\label{fig:GTCWhiteLC}}
\end{figure}

The light-curve data were modeled by the analytic transit model \citep{2002ApJ...580L.171M} together with the Gaussian process \citep[GP;][]{2006gpml.book.....R} to account for the systematics trends. The transit model was assumed as the mean function of GP, and implemented using the Python package \texttt{batman} \citep{2015PASP..127.1161K}, where the quadratic limb darkening law was adopted. The limb darkening coefficients (LDC) were calculated using the Kurucz ATLAS9 stellar atmosphere models with stellar effective temperature $T_\mathrm{eff}=5750$~K, surface gravity $\log g_\star=3.9$, and metallicity $\mathrm{[Fe/H]}=-0.18$ \citep{2015MNRAS.450.1879E}. The two LDC were always fitted with Gaussian priors of the width $\sigma_\mathrm{LDC}=0.1$ to take advantage of the stellar physical information. The GP was implemented using the Python package \texttt{george} \citep{2015ITPAM..38..252A}, with the correlated systematics modeled by the covariance matrix in the form of the squared exponential (SE) kernel,
\begin{align}
k_\mathrm{SE}(x_i,x_j,\theta)&=A^2\exp\Bigg[-\sum\limits_{\alpha=1}^{N}\bigg(\frac{x_{\alpha,i}-x_{\alpha,j}}{L_\alpha}\bigg)^2\Bigg]. 
\end{align}
The GP input variables $x_{i,j}$ could be time or auxiliary trends such as position drifts ($x$, $y$) and seeing variations ($s_x$, $s_y$) measured in the cross-dispersion and spatial directions. Uniform priors were used for all the GP hyperparameters. When time is used as a GP input, the corresponding scale hyperparameter $L_t$ is always forced to be no shorter than the ingress/egress time (0.01749 days in this case). The posterior distributions were determined by the Python package \texttt{emcee} \citep{2013PASP..125..306F} with the Affine Invariant Markov chain Monte Carlo (MCMC) Ensemble sampler. 

We chose to jointly analyze our GTC/OSIRIS light curves together with the NOT/ALFOSC light curves published by \citet{2017A&A...602L..15P}. For the white-color light curves, the fitted parameters were inclination $i$, scaled semimajor axis \aRs, mid-transit time \Tmid, planet-to-star radius ratio \RpRs, LDC $u_1$ and $u_2$, and the GP hyperparameters. Period $P$ was fixed to the literature value. Eccentricity $e$ was fixed to zero. In this joint modeling, the white-color light curves of the two transits shared the same $i$ and \aRs, but were allowed to have different \Tmid, \RpRs, $u_1$, and $u_2$. The NOT/ALFOSC data adopted time and seeing ($s_y$, measured as the FWHM of the spatial profile) as the GP inputs, while the GTC/OSIRIS data adopted time, seeing $s_y$, and cross-dispersion drift $\Delta x$ as the GP inputs. Our MCMC process for the white-color light-curve joint analysis consisted of 90 walkers, each with two burn-ins of 500 steps and another 3000 steps.

The spectroscopic light curves were modeled individually for each transit. We first constructed the empirical common-mode noise model by dividing the white-color light-curve data by the best-fitting analytic transit model. Every spectroscopic light curve was divided by this common-mode noise. The modeling of the corrected spectroscopic light curves adopted the same GP inputs as their white-color light curves. The fitted parameters were \RpRs, $u_1$, $u_2$, and the GP hyperparameters. The other parameters, including $i$, \aRs, and \Tmid, were fixed to the best-fitting values derived from the white-color light curve of corresponding transit. Our MCMC process for the spectroscopic light curves consisted of 32 walkers, each with two burn-ins of 500 steps and another 2500 steps.

To combine the two data sets, a corrective constant offset of $\Delta\RpRs=0.004056$ was subtracted from the NOT/ALFOSC transmission spectrum. This was determined in the common wavelength range 525--590~nm between the NOT/ALFOSC and GTC/OSIRIS stellar spectra. The wavelength was limited up to 590~nm to avoid any imperfect correction of second-order contamination in the NOT/ALFOSC stellar spectra. Light curves of this 525--590~nm band were created after the difference in instrumental response function was corrected. A joint modeling of the NOT/ALFOSC and GTC/OSIRIS 525--590~nm light curves was performed in a similar manner to the spectroscopic light curves, except that they shared the same $u_1$ and $u_2$. We obtained $\RpRs=0.10472\pm0.0024$ for NOT/ALFOSC, $\RpRs=0.10066\pm0.0017$ for GTC/OSIRIS, and $\Delta\RpRs=0.004056\pm0.0029$ as the difference. Such an offset could come from an instrumental bias within the 1$\sigma$ uncertainty or  variation of stellar flux baseline caused by stellar activity. Given the consistency in the spectral shapes between the NOT/ALFOSC and GTC/OSIRIS transmission spectra, it is not likely caused by the stellar activity.

\section{Results and discussion}
\label{sec:discuss}

We present the derived parameters from the white-color light-curve joint analysis in Table \ref{tab:gtc_param}. In addition to having refined the transit parameters $i$ and \aRs, we also have revised the orbital ephemeris by combining our two mid-transit times with that in the discovery paper. We show the GTC/OSIRIS white-color light curve in Fig.~\ref{fig:GTCWhiteLC} and present the spectroscopic light curves and their transit depths in Appendix \ref{app:addtabfig}.

\begin{table}
     \small
     \centering
     \caption{Derived parameters from the GTC/OSIRIS and NOT/ALFOSC joint analysis}
     \label{tab:gtc_param}
     \begin{tabular}{rccc}
     \hline\hline\noalign{\smallskip}
     Parameter & GTC/OSIRIS & NOT/ALFOSC\\\noalign{\smallskip}
     \hline\noalign{\smallskip}
     \multicolumn{3}{c}{\it Transit parameters\dotfill}\\\noalign{\smallskip}
     $R_{\rm p}/R_\star$ & 0.09992 $\pm$ 0.0018 & 0.10550 $\pm$ 0.0016 \\\noalign{\smallskip}
        $i$ [$^{\circ}$] & \multicolumn{2}{c}{87.88 $\pm$ 0.32} \\\noalign{\smallskip}
             $a/R_\star$ & \multicolumn{2}{c}{7.846 $\pm$ 0.089} \\\noalign{\smallskip}
                   $u_1$ & 0.237 $\pm$ 0.068 & 0.434 $\pm$ 0.053 \\\noalign{\smallskip}
                   $u_2$ & 0.221 $\pm$ 0.079 & 0.209 $\pm$ 0.076 \\\noalign{\smallskip}
     \hline\noalign{\smallskip}
     \multicolumn{3}{c}{\it Mid-transit times\dotfill}\\\noalign{\smallskip}
     $T_{\rm mid}$ [MJD\tablefootmark{a}] & 8138.670144 $\pm$ 0.00030 & 7808.602736 $\pm$ 0.00022 \\\noalign{\smallskip}
     \hline\noalign{\smallskip}
     \multicolumn{3}{c}{\it Revised ephemeris\dotfill}\\\noalign{\smallskip}
     $T_0$ [MJD\tablefootmark{a}] & \multicolumn{2}{c}{7248.741276 $\pm$ 0.000068} \\\noalign{\smallskip}
     $P$ [days]   & \multicolumn{2}{c}{4.17807015 $\pm$ 0.00000057} \\\noalign{\smallskip}
    \hline\noalign{\smallskip}
    \end{tabular}
    \tablefoot{\tablefoottext{a}{$\mathrm{MJD}=\mathrm{BJD}_\mathrm{TDB}-2450000$.}}
\end{table}

\begin{figure*}
\centering
\includegraphics[width=1.0\linewidth]{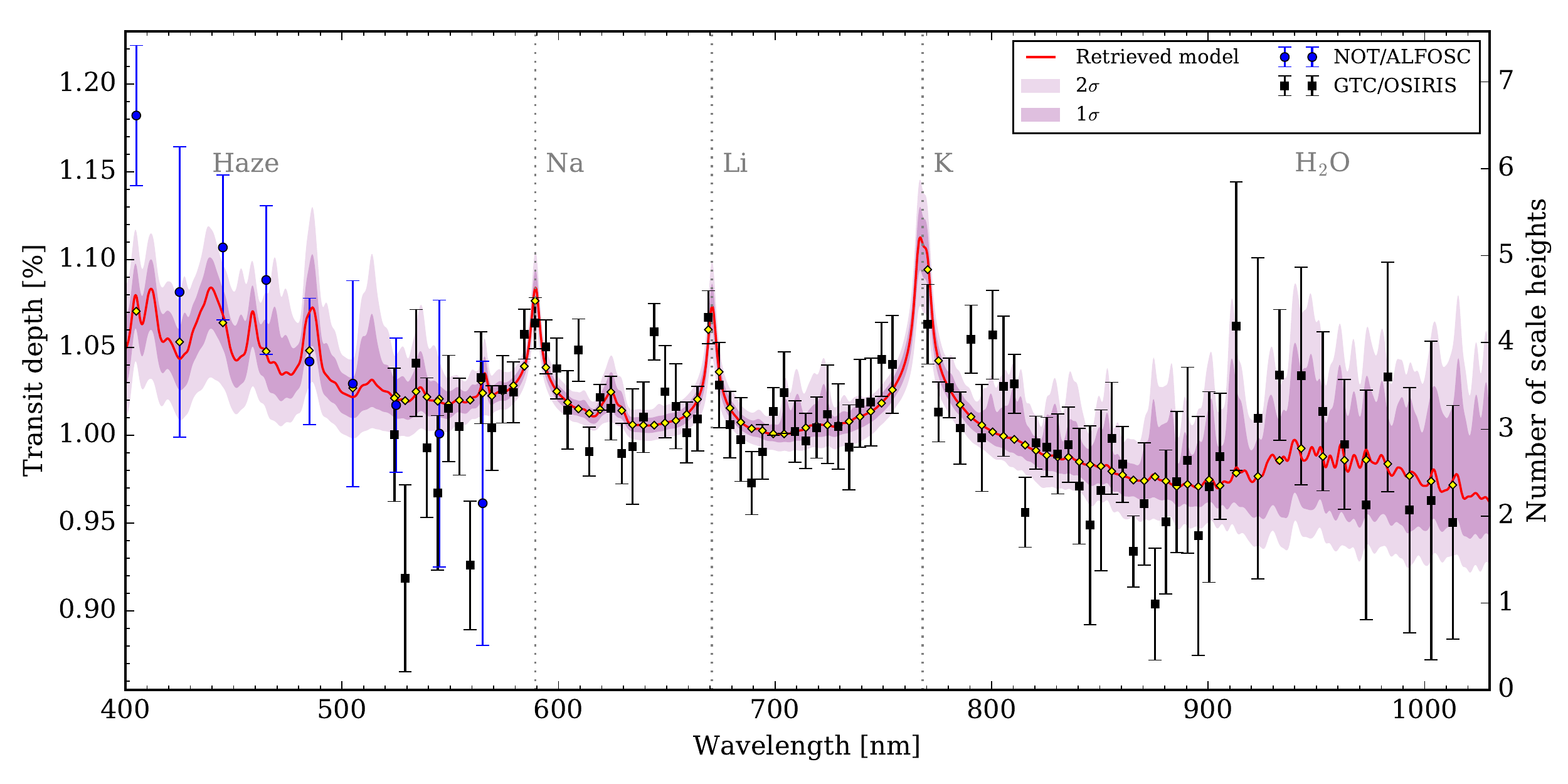}
\caption{Transmission spectrum of \object{WASP-127b} and retrieved models. The blue circles and black squares with error bars are the observed spectrum by NOT/ALFOSC and GTC/OSIRIS, respectively. This spectrum shows an enhanced slope at the blue optical, strong absorption peaks at 589.3~nm, 670.8~nm, and 768.2~nm, and another bump at the red optical. These features can be explained by the model spectrum when including opacities resulting from haze, Na, Li, K, and H$_2$O, respectively. The red curve shows the retrieved median model while the shaded areas show the 1$\sigma$ and 2$\sigma$ confidence regions. The yellow diamonds show the binned version for the retrieved median model. }
\label{fig:spectrum}
\end{figure*}

\subsection{Transmission spectrum}
\label{sec:transpec}

The GTC/OSIRIS and NOT/ALFOSC transmission spectra agree well with each other in the common wavelength range of 520--945~nm (see Fig.~\ref{fig:GTCTranSpec}), although the latter has a lower spectral resolution and larger uncertainties. The large scatter of NOT/ALFOSC data might be attributed to the imperfect correction of the second-order contamination \citep{2007AN....328..948S,2017A&A...602L..15P}. Given that the GTC/OSIRIS data have sufficient quality at a higher spectral resolution, we chose not to include the NOT/ALFOSC data at wavelengths longer than 590~nm in the subsequent analysis and discussion. We verified that including the extra NOT/ALFOSC data ($\lambda>590$~nm) does not change our subsequent results, but introduces more noise.

The combined transmission spectrum spans 5.6 atmospheric scale heights ($H/R_\star=0.00243$) from maximum to minimum transit depths. It is inconsistent with a constant flat line at 5.6$\sigma$ level, and the cloud-free 1400~K solar composition fiducial atmospheric model \citep{2017PASP..129d4402K} at 10.8$\sigma$ level. However, the spectrum does exhibit four distinct broad spectral features, which can be well explained by the linearly scaled fiducial model individually. At the blue wavelengths ($\lambda<580$~nm), the combined transmission spectrum shows a slope
\begin{align}
m_\mathrm{obs}=\frac{dR_\mathrm{p}}{d\ln\lambda}=(-0.0175 \pm 0.0040)R_\star. 
\end{align}
This can be converted to a dimensionless slope $\mathcal{S}=m_\mathrm{obs}/H_\mathrm{eq}=-7.2\pm 1.6$, which is consistent with hazes composed of sulphide (e.g., MnS and ZnS) with particle sizes less than 0.1~$\mu$m \citep{2017MNRAS.471.4355P}. Focusing on the GTC/OSIRIS measurements, the transmission spectrum shows the pressure-broadened wings of Na and K centered at around 589.3~nm and 768.2~nm, respectively. At red wavelengths ($\lambda>900$~nm), the transmission spectrum starts to bump and then drop, which is consistent with the water absorption feature. 

\subsection{Atmospheric retrieval}
\label{sec:retrieval}

To globally retrieve the atmospheric properties at the day-night terminator region, we performed a spectral retrieval modeling on the combined transmission spectrum of \object{WASP-127b}. We used an atmospheric retrieval code for transmission spectra adapted from recent works \citep{2018MNRAS.474..271G} and used the cloud/haze parameterization of \citet{2017MNRAS.469.1979M}. The haze is included as $\sigma=a\sigma_0(\lambda/\lambda_0)^\gamma$, where $\gamma$ is the scattering slope, $a$ is the Rayleigh-enhancement factor, and $\sigma_0$ is the H$_2$ Rayleigh scattering cross section ($5.31\times10^{-31}$~m$^2$) at the reference wavelength $\lambda_0=350$~nm. The model computes line-by-line radiative transfer in a transmission geometry, assuming a plane parallel planetary atmosphere in hydrostatic equilibrium and local thermodynamic equilibrium. The model also assumes the reference pressure as a free parameter, which is the pressure at an assumed radius $R_\mathrm{p}=1.37$~$R_\mathrm{Jup}$ \citep{2017A&A...599A...3L}. Considering $R_\mathrm{p}$ as a free parameter does not significantly change our results. The chemical composition and temperature profile are free parameters in the model. We performed a set of retrievals considering the molecules, metal oxides and hydrides, and atomic species that could be present in hot Jupiter atmospheres: i.e., H$_2$O, CO, CH$_4$, NH$_3$, CO$_2$, TiO, AlO, FeH, TiH, CrH, Na, K, Li, V, and Fe \citep{2016SSRv..205..285M}. Our atmospheric pressure-temperature (P-T) model consists of six parameters \citep{2009ApJ...707...24M} and we consider models in the range from clear to cloudy, both with and without scattering hazes and with inhomogeneous cloud coverage. The opacities for the chemical species are adopted from \citet{2017MNRAS.472.2334G,2018MNRAS.474..271G}. The Bayesian inference and parameter estimation is conducted using the nested sampling algorithm implemented via the MultiNest application \citep{2009MNRAS.398.1601F}, as pursued in previous studies \citep{2017MNRAS.469.1979M,2018MNRAS.474..271G}.  

The optical transmission spectrum of \object{WASP-127b} provides strong constraints on its atmospheric composition. We report the detection of K at a confidence level of 5.0$\sigma$, Na at 4.1$\sigma$, and Li at 3.4$\sigma$ in the spectrum along with an indication of H$_2$O (see Table \ref{tab:bf_retrieval} for the Bayesian model comparison). Figure \ref{fig:spectrum} shows the best-fit spectrum to the data along with the significance contours. The models without Na, K, or Li fail to explain the peaks in absorption at $\sim$589.3~nm, $\sim$670.8~nm, and $\sim$768.2~nm. We do not have statistically significant detections of any other chemical species considered in the model. 

The degeneracy between reference planet radius, reference pressure, and chemical abundances can make it difficult to accurately retrieve chemical abundances in uniformly cloudy atmospheres \citep{2012ApJ...753..100B,2014RSPTA.37230086G,2017MNRAS.470.2972H}. However, this degeneracy can be broken or reduced if the atmosphere is cloud-free \citep{2014RSPTA.37230086G,2017MNRAS.470.2972H} or partially cloudy \citep{2017MNRAS.469.1979M}, where H$_2$ Rayleigh scattering or pressure-broadened line wings can help constrain the reference pressure level. Our partially cloudy model enables us to reduce the impact of this degeneracy and account for any remaining correlation in the derived uncertainties. In particular, the presence of optical spectra helps further mitigate the degeneracy, especially when the spectrum is not flat. 

Our modeling retrieved volume mixing ratios of $\log(X_\mathrm{Na})= -3.17 ^{+1.03}_{-1.46}$, $\log(X_\mathrm{K})=-2.13 ^{+0.85}_{-1.32}$, and $\log(X_\mathrm{Li})= -3.17 ^{+0.97}_{-1.51}$ for Na, K, and Li, respectively. The retrieved Na abundance is tentatively super-solar \citep{2009ARA&A..47..481A} at 1.8$\sigma$, while the abundances of K and Li are super-solar at 3.7$\sigma$ and 5.1$\sigma$, respectively. The Li abundance $\log(A_\mathrm{Li})=\log(X_\mathrm{Li})+12$ is also significantly higher than the super-solar value of the host star \citep[$1.97\pm0.09$; ][]{2017A&A...599A...3L}. The detection of Li in the planet atmosphere could open a new window to understand lithium depletion in planet-host stars and planet formation history \citep[e.g.,][]{2008A&A...489L..53B,2009Natur.462..189I,2010A&A...521A..44B}. We have verified that the retrieved volume mixing ratios remain well consistent within 1$\sigma$ error bar even if the reference planet radius is a free parameter instead of assumed as $R_\mathrm{p}=1.37$~$R_\mathrm{Jup}$ (see Appendix~\ref{app:retrieval_tests} for retrieval tests). The retrieved haze is $\sim$8500--250000 (68.3\% confidence interval) stronger than H$_2$ Rayleigh scattering, has a coverage of $\phi= 52^{+10}_{-9}$\%, and a power-law exponent of $\gamma=-7.36 ^{+2.33}_{-2.56}$. The P-T profile is relatively unconstrained by the data (see Fig.~\ref{fig:ptprof}). 

Our models also considered the presence of water in the atmosphere of the planet. Although the spectral shape in the wavelength range 833--1018~nm resembles a water absorption owing to low flux and fringing effect, the current error in the data within 900-1018~nm is relatively large and does not constrain the abundance of water in the global atmospheric retrieval. We find a nominal water signature with a relatively weak abundance constraint of $\log(\mathrm{H_2O})=-2.60 ^{+0.94}_{-4.56}$, which can be confirmed with HST near-infrared spectroscopy in the near term, and with James Webb Space Telescope (JWST) in the future.

\section{Conclusions}\label{sec:conclusions}

We have observed one transit of the super-Neptune \object{WASP-127b} using the long-slit mode of GTC/OSIRIS. We revised the transit parameters by jointly analyzing the GTC/OSIRIS white-color light curve with the already published NOT/ALFOSC light curve using Gaussian processes. The resulting transmission spectra from the two transits are consistent in spectral shape. With the combined transmission spectrum, we detected a scattering haze at the blue wavelengths, the pressure-broadened spectral profiles of Na, K and Li absorption, and found a hint of water absorption at the red wavelengths. We inferred a tentatively super-solar abundance for Na, significantly super-solar abundances for K and Li, a coverage of $\sim$52\% for haze, and a weakly constrained abundance for water based on the spectral retrieval modeling. Our results showcase that large-aperture ground-based telescopes could result in high-quality spectroscopy that is comparable to or even better than what HST can do \citep[also see][]{2017Natur.549..238S}. Thanks to its rare physical parameters and the confirmation of relatively clear sky, \object{WASP-127b} will become a benchmark for exoplanet atmospheric characterization in the era of JWST and high-resolution spectroscopy at 30~m class telescopes.

\begin{acknowledgements}
    This research is based on observations made with the Gran Telescopio Canarias (GTC), installed in the Spanish Observatorio del Roque de los Muchachos, operated on the island of La Palma by the Instituto de Astrof\'isica de Canarias.  
    This work is partly financed by the Spanish Ministry of Economics and Competitiveness through grant ESP2013-48391-C4-2-R. 
    G.C. acknowledges the support by the National Natural Science Foundation of China (Grant No. 11503088) and the Natural Science Foundation of Jiangsu Province (Grant No. BK20151051), and the Minor Planet Foundation of the Purple Mountain Observatory.
    This research has made use of Matplotlib \citep{2007CSE.....9...90H} and the VizieR catalog access tool, CDS, Strasbourg, France \citep{2000A&AS..143...23O}. 
    The authors thank the anonymous referee for useful comments on the manuscript.
\end{acknowledgements}

\bibliographystyle{aa} 
\bibliography{ref_db} 

\begin{thebibliography}{41}
\expandafter\ifx\csname natexlab\endcsname\relax\def\natexlab#1{#1}\fi

\bibitem[{{Ambikasaran} {et~al.}(2015){Ambikasaran}, {Foreman-Mackey},
  {Greengard}, {Hogg}, \& {O'Neil}}]{2015ITPAM..38..252A}
{Ambikasaran}, S., {Foreman-Mackey}, D., {Greengard}, L., {Hogg}, D.~W., \&
  {O'Neil}, M. 2015, IEEE Transactions on Pattern Analysis and Machine
  Intelligence, 38 [\eprint[arXiv]{1403.6015}]

\bibitem[{{Arcangeli} {et~al.}(2018){Arcangeli}, {D{\'e}sert}, {Line}, {Bean},
  {Parmentier}, {Stevenson}, {Kreidberg}, {Fortney}, {Mansfield}, \&
  {Showman}}]{2018ApJ...855L..30A}
{Arcangeli}, J., {D{\'e}sert}, J.-M., {Line}, M.~R., {et~al.} 2018, \apjl, 855,
  L30

\bibitem[{{Asplund} {et~al.}(2009){Asplund}, {Grevesse}, {Sauval}, \&
  {Scott}}]{2009ARA&A..47..481A}
{Asplund}, M., {Grevesse}, N., {Sauval}, A.~J., \& {Scott}, P. 2009, \araa, 47,
  481

\bibitem[{{Baraffe} \& {Chabrier}(2010)}]{2010A&A...521A..44B}
{Baraffe}, I. \& {Chabrier}, G. 2010, \aap, 521, A44

\bibitem[{{Benneke} \& {Seager}(2012)}]{2012ApJ...753..100B}
{Benneke}, B. \& {Seager}, S. 2012, \apj, 753, 100

\bibitem[{{Bouvier}(2008)}]{2008A&A...489L..53B}
{Bouvier}, J. 2008, \aap, 489, L53

\bibitem[{{Chen} {et~al.}(2017{\natexlab{a}}){Chen}, {Guenther}, {Pall{\'e}},
  {Nortmann}, {Nowak}, {Kunz}, {Parviainen}, \& {Murgas}}]{2017A&A...600A.138C}
{Chen}, G., {Guenther}, E.~W., {Pall{\'e}}, E., {et~al.} 2017{\natexlab{a}},
  \aap, 600, A138

\bibitem[{{Chen} {et~al.}(2017{\natexlab{b}}){Chen}, {Pall{\'e}}, {Nortmann},
  {Murgas}, {Parviainen}, \& {Nowak}}]{2017A&A...600L..11C}
{Chen}, G., {Pall{\'e}}, E., {Nortmann}, L., {et~al.} 2017{\natexlab{b}}, \aap,
  600, L11

\bibitem[{{Eastman} {et~al.}(2010){Eastman}, {Siverd}, \&
  {Gaudi}}]{2010PASP..122..935E}
{Eastman}, J., {Siverd}, R., \& {Gaudi}, B.~S. 2010, \pasp, 122, 935

\bibitem[{{Espinoza} \& {Jord{\'a}n}(2015)}]{2015MNRAS.450.1879E}
{Espinoza}, N. \& {Jord{\'a}n}, A. 2015, \mnras, 450, 1879

\bibitem[{{Feroz} {et~al.}(2009){Feroz}, {Hobson}, \&
  {Bridges}}]{2009MNRAS.398.1601F}
{Feroz}, F., {Hobson}, M.~P., \& {Bridges}, M. 2009, \mnras, 398, 1601

\bibitem[{{Foreman-Mackey} {et~al.}(2013){Foreman-Mackey}, {Hogg}, {Lang}, \&
  {Goodman}}]{2013PASP..125..306F}
{Foreman-Mackey}, D., {Hogg}, D.~W., {Lang}, D., \& {Goodman}, J. 2013, \pasp,
  125, 306

\bibitem[{{Gandhi} \& {Madhusudhan}(2017)}]{2017MNRAS.472.2334G}
{Gandhi}, S. \& {Madhusudhan}, N. 2017, \mnras, 472, 2334

\bibitem[{{Gandhi} \& {Madhusudhan}(2018)}]{2018MNRAS.474..271G}
{Gandhi}, S. \& {Madhusudhan}, N. 2018, \mnras, 474, 271

\bibitem[{{Griffith}(2014)}]{2014RSPTA.37230086G}
{Griffith}, C.~A. 2014, Philosophical Transactions of the Royal Society of
  London Series A, 372, 20130086

\bibitem[{{Heng} \& {Kitzmann}(2017)}]{2017MNRAS.470.2972H}
{Heng}, K. \& {Kitzmann}, D. 2017, \mnras, 470, 2972

\bibitem[{{Horne}(1986)}]{1986PASP...98..609H}
{Horne}, K. 1986, \pasp, 98, 609

\bibitem[{{Hunter}(2007)}]{2007CSE.....9...90H}
{Hunter}, J.~D. 2007, Computing in Science and Engineering, 9, 90

\bibitem[{{Israelian} {et~al.}(2009){Israelian}, {Delgado Mena}, {Santos},
  {Sousa}, {Mayor}, {Udry}, {Dom{\'{\i}}nguez Cerde{\~n}a}, {Rebolo}, \&
  {Randich}}]{2009Natur.462..189I}
{Israelian}, G., {Delgado Mena}, E., {Santos}, N.~C., {et~al.} 2009, \nat, 462,
  189

\bibitem[{{Kempton} {et~al.}(2017){Kempton}, {Lupu}, {Owusu-Asare}, {Slough},
  \& {Cale}}]{2017PASP..129d4402K}
{Kempton}, E.~M.-R., {Lupu}, R., {Owusu-Asare}, A., {Slough}, P., \& {Cale}, B.
  2017, \pasp, 129, 044402

\bibitem[{{Kreidberg}(2015)}]{2015PASP..127.1161K}
{Kreidberg}, L. 2015, \pasp, 127, 1161

\bibitem[{{Kreidberg} {et~al.}(2014){Kreidberg}, {Bean}, {D{\'e}sert}, {Line},
  {Fortney}, {Madhusudhan}, {Stevenson}, {Showman}, {Charbonneau},
  {McCullough}, {Seager}, {Burrows}, {Henry}, {Williamson}, {Kataria}, \&
  {Homeier}}]{2014ApJ...793L..27K}
{Kreidberg}, L., {Bean}, J.~L., {D{\'e}sert}, J.-M., {et~al.} 2014, \apjl, 793,
  L27

\bibitem[{{Lam} {et~al.}(2017){Lam}, {Faedi}, {Brown}, {Anderson}, {Delrez},
  {Gillon}, {H{\'e}brard}, {Lendl}, {Mancini}, {Southworth}, {Smalley},
  {Triaud}, {Turner}, {Hay}, {Armstrong}, {Barros}, {Bonomo}, {Bouchy},
  {Boumis}, {Collier Cameron}, {Doyle}, {Hellier}, {Henning}, {Jehin}, {King},
  {Kirk}, {Louden}, {Maxted}, {McCormac}, {Osborn}, {Palle}, {Pepe},
  {Pollacco}, {Prieto-Arranz}, {Queloz}, {Rey}, {S{\'e}gransan}, {Udry},
  {Walker}, {West}, \& {Wheatley}}]{2017A&A...599A...3L}
{Lam}, K.~W.~F., {Faedi}, F., {Brown}, D.~J.~A., {et~al.} 2017, \aap, 599, A3

\bibitem[{{MacDonald} \& {Madhusudhan}(2017)}]{2017MNRAS.469.1979M}
{MacDonald}, R.~J. \& {Madhusudhan}, N. 2017, \mnras, 469, 1979

\bibitem[{{Madhusudhan} {et~al.}(2016){Madhusudhan}, {Ag{\'u}ndez}, {Moses}, \&
  {Hu}}]{2016SSRv..205..285M}
{Madhusudhan}, N., {Ag{\'u}ndez}, M., {Moses}, J.~I., \& {Hu}, Y. 2016, \ssr,
  205, 285

\bibitem[{{Madhusudhan} {et~al.}(2014){Madhusudhan}, {Amin}, \&
  {Kennedy}}]{2014ApJ...794L..12M}
{Madhusudhan}, N., {Amin}, M.~A., \& {Kennedy}, G.~M. 2014, \apjl, 794, L12

\bibitem[{{Madhusudhan} \& {Seager}(2009)}]{2009ApJ...707...24M}
{Madhusudhan}, N. \& {Seager}, S. 2009, \apj, 707, 24

\bibitem[{{Mandel} \& {Agol}(2002)}]{2002ApJ...580L.171M}
{Mandel}, K. \& {Agol}, E. 2002, \apjl, 580, L171

\bibitem[{{Mazeh} {et~al.}(2016){Mazeh}, {Holczer}, \&
  {Faigler}}]{2016A&A...589A..75M}
{Mazeh}, T., {Holczer}, T., \& {Faigler}, S. 2016, \aap, 589, A75

\bibitem[{{Mordasini} {et~al.}(2016){Mordasini}, {van Boekel}, {Molli{\`e}re},
  {Henning}, \& {Benneke}}]{2016ApJ...832...41M}
{Mordasini}, C., {van Boekel}, R., {Molli{\`e}re}, P., {Henning}, T., \&
  {Benneke}, B. 2016, \apj, 832, 41

\bibitem[{{Nikolov} {et~al.}(2018){Nikolov}, {Sing}, {Fortney}, {Goyal},
  {Drummond}, {Evans}, {Gibson}, {De Mooij}, {Rustamkulov}, {Wakeford},
  {Smalley}, {Burgasser}, {Hellier}, {Helling}, {Mayne}, {Madhusudhan},
  {Kataria}, {Baines}, {Carter}, {Ballester}, {Barstow}, {McCleery}, \&
  {Spake}}]{Nikolov2018bNaturePaperWASP96b}
{Nikolov}, N., {Sing}, D.~K., {Fortney}, J.~J., {et~al.} 2018, \nat, 000, 00

\bibitem[{{{\"O}berg} {et~al.}(2011){{\"O}berg}, {Murray-Clay}, \&
  {Bergin}}]{2011ApJ...743L..16O}
{{\"O}berg}, K.~I., {Murray-Clay}, R., \& {Bergin}, E.~A. 2011, \apjl, 743, L16

\bibitem[{{Ochsenbein} {et~al.}(2000){Ochsenbein}, {Bauer}, \&
  {Marcout}}]{2000A&AS..143...23O}
{Ochsenbein}, F., {Bauer}, P., \& {Marcout}, J. 2000, \aaps, 143, 23

\bibitem[{{Palle} {et~al.}(2017){Palle}, {Chen}, {Prieto-Arranz}, {Nowak},
  {Murgas}, {Nortmann}, {Pollacco}, {Lam}, {Montanes-Rodriguez}, {Parviainen},
  \& {Casasayas-Barris}}]{2017A&A...602L..15P}
{Palle}, E., {Chen}, G., {Prieto-Arranz}, J., {et~al.} 2017, \aap, 602, L15

\bibitem[{{Pinhas} \& {Madhusudhan}(2017)}]{2017MNRAS.471.4355P}
{Pinhas}, A. \& {Madhusudhan}, N. 2017, \mnras, 471, 4355

\bibitem[{{Rasmussen} \& {Williams}(2006)}]{2006gpml.book.....R}
{Rasmussen}, C.~E. \& {Williams}, C.~K.~I. 2006, {Gaussian Processes for
  Machine Learning}

\bibitem[{{S{\'a}nchez} {et~al.}(2012){S{\'a}nchez}, {Aguiar-Gonz{\'a}lez},
  {Barreto}, {Becerril}, {Bland-Hawthorn}, {Bongiovanni}, {Cepa}, {Correa},
  {Chapa}, {Ederoclite}, {Espejo}, {Farah}, {Fragoso}, {Fern{\'a}ndez},
  {Flores}, {Fuentes}, {Gago}, {Garfias}, {Gigante}, {Gonz{\'a}lez},
  {Gonz{\'a}lez-Escalera}, {Hern{\'a}ndez}, {Hernandez}, {Herrera}, {Herrera},
  {Joven}, {Langarica}, {Lara}, {L{\'o}pez}, {L{\'o}pez}, {Militellon},
  {Moreno}, {Peraza}, {P{\'e}rez}, {P{\'e}rez}, {Rasilla}, {Rosich}, {Tejada},
  {Tinoco}, {Vaz}, \& {Villegas}}]{2012SPIE.8446E..4TS}
{S{\'a}nchez}, B., {Aguiar-Gonz{\'a}lez}, M., {Barreto}, R., {et~al.} 2012, in
  Society of Photo-Optical Instrumentation Engineers (SPIE) Conference Series,
  Vol. 8446, Society of Photo-Optical Instrumentation Engineers (SPIE)
  Conference Series, 4

\bibitem[{{Sedaghati} {et~al.}(2017){Sedaghati}, {Boffin}, {MacDonald},
  {Gandhi}, {Madhusudhan}, {Gibson}, {Oshagh}, {Claret}, \&
  {Rauer}}]{2017Natur.549..238S}
{Sedaghati}, E., {Boffin}, H.~M.~J., {MacDonald}, R.~J., {et~al.} 2017, \nat,
  549, 238

\bibitem[{{Sing} {et~al.}(2016){Sing}, {Fortney}, {Nikolov}, {Wakeford},
  {Kataria}, {Evans}, {Aigrain}, {Ballester}, {Burrows}, {Deming},
  {D{\'e}sert}, {Gibson}, {Henry}, {Huitson}, {Knutson}, {Etangs}, {Pont},
  {Showman}, {Vidal-Madjar}, {Williamson}, \& {Wilson}}]{2016Natur.529...59S}
{Sing}, D.~K., {Fortney}, J.~J., {Nikolov}, N., {et~al.} 2016, \nat, 529, 59

\bibitem[{{Stanishev}(2007)}]{2007AN....328..948S}
{Stanishev}, V. 2007, Astronomische Nachrichten, 328, 948

\bibitem[{{Wakeford} {et~al.}(2017){Wakeford}, {Sing}, {Kataria}, {Deming},
  {Nikolov}, {Lopez}, {Tremblin}, {Amundsen}, {Lewis}, {Mandell}, {Fortney},
  {Knutson}, {Benneke}, \& {Evans}}]{2017Sci...356..628W}
{Wakeford}, H.~R., {Sing}, D.~K., {Kataria}, T., {et~al.} 2017, Science, 356,
  628

\end{thebibliography}


\begin{appendix}

\section{Spectral movements caused by guiding problems}\label{app:guiding}

\begin{figure}[h!]
\centering
\includegraphics[width=1\linewidth]{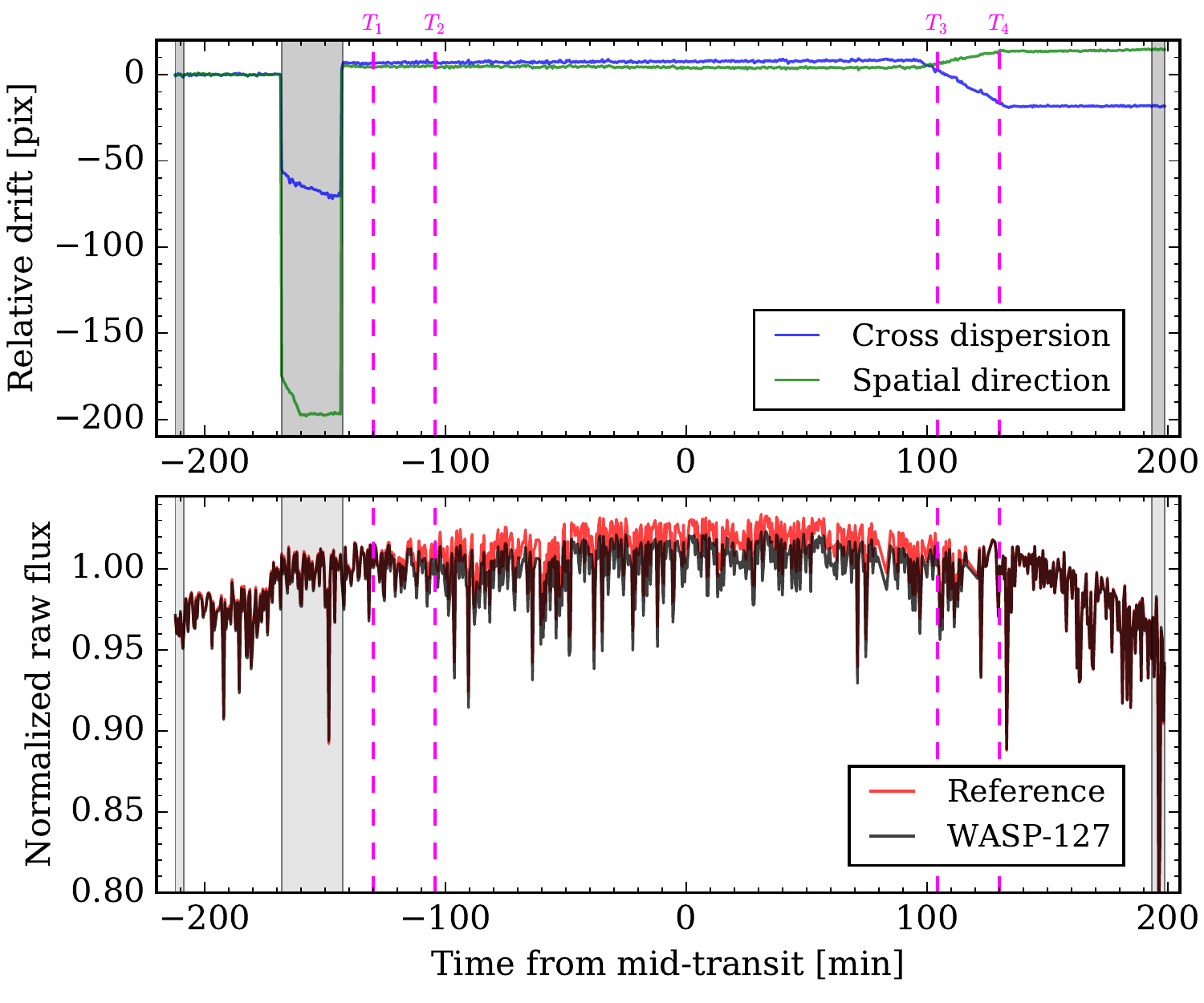}
\caption{Spectral movements and the raw flux time series. The top panel shows the drift of spectra location in the cross-dispersion (blue) and spatial (green) directions. The bottom panel shows the normalized raw flux of \object{WASP-127} (black) and its reference star (red). The gray shaded areas indicate the discarded exposures that are not included in subsequent light-curve modeling, where the two regions on the edge are impacted by high airmass, the region on the right edge is also impacted by morning twilight, and the middle region has severely lost guiding. The vertical dashed lines indicate the first ($T_1$), second ($T_2$), third ($T_3$), and fourth ($T_4$) contact of the transit event.\label{fig:GTCTrends}}
\end{figure}

Guiding problems occurred twice during the observation. At 01:10 UT, the target jumped to a location $\sim$180 pixels away in the spatial direction. The target was drifting for another 20 pixels until it was put back to a location that is close to the original location. It lost guiding again at around 06:12 UT with slow drifting for $\sim$10 pixels as well. The inspection of absorption lines in the acquired spectra shows that such a drift and jump also exist in the cross-dispersion direction. The absorption lines jumped $\sim$60 pixels for the first jump and the position was $\sim$7 pixels away from the original location when it was put back. The second jump caused the absorption lines to move $\sim$25 pixels away. Figure \ref{fig:GTCTrends} shows the drift of spectra location on the CCD in both cross-dispersion and spatial directions, and the raw flux sequence of the two stars. 

\section{Additional retrieval test on the transmission spectrum}\label{app:retrieval_tests}

\begin{figure}[h!]
\centering
\includegraphics[width=0.9\linewidth]{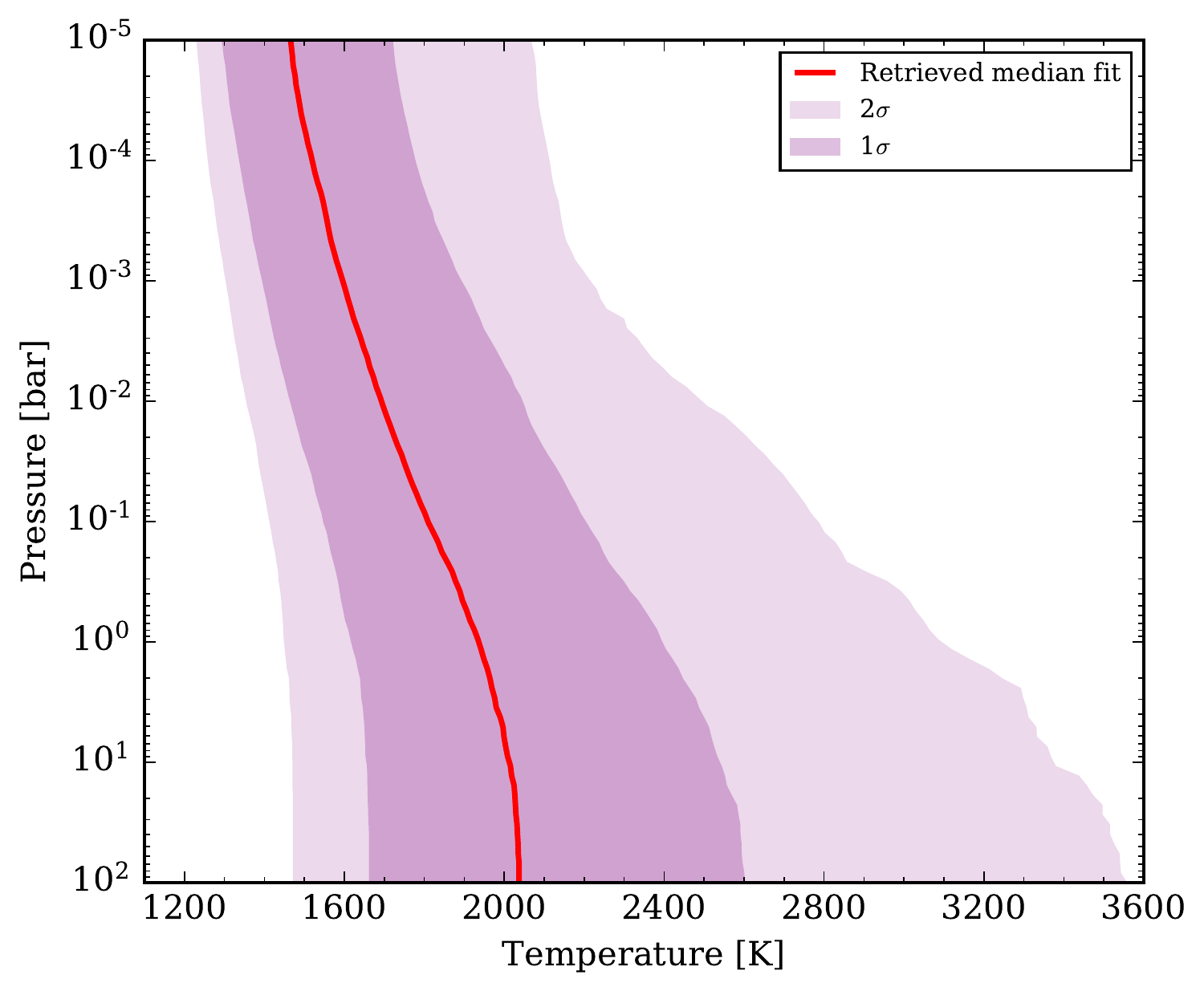}
\caption{Retrieved pressure temperature profile. The shaded areas show the 1$\sigma$ and 2$\sigma$ confidence regions.}
\label{fig:ptprof}
\end{figure}

\begin{figure*}[h!]
\centering
\includegraphics[width=\linewidth]{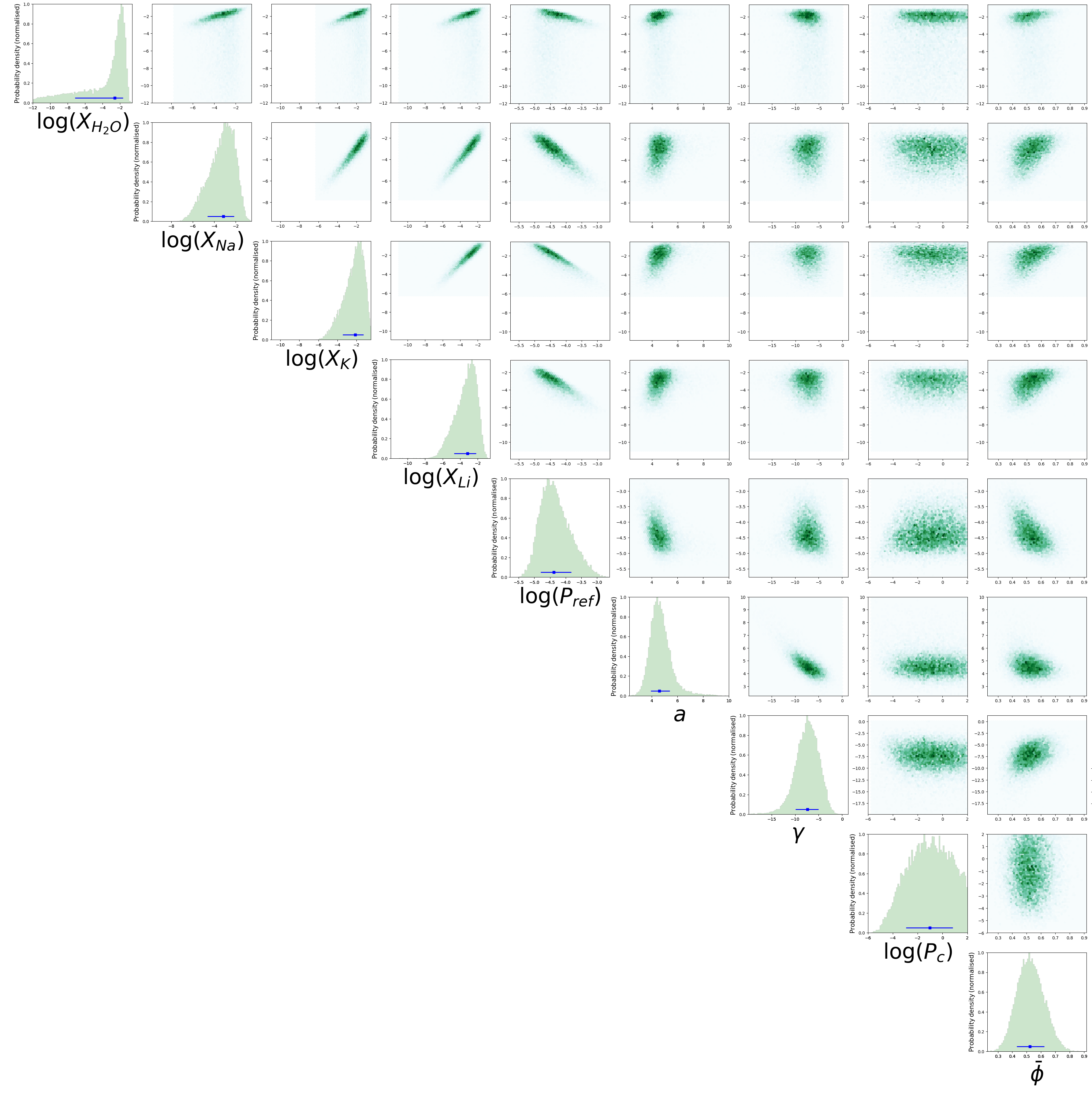}
\caption{Marginalized posterior probability densities for the retrieved species and haze parameters with the reference planet radius fixed at $R_\mathrm{p}=1.37$~$R_\mathrm{Jup}$. }
\label{fig:posterior}
\end{figure*}

\begin{figure*}[h!]
\centering
\includegraphics[width=\linewidth]{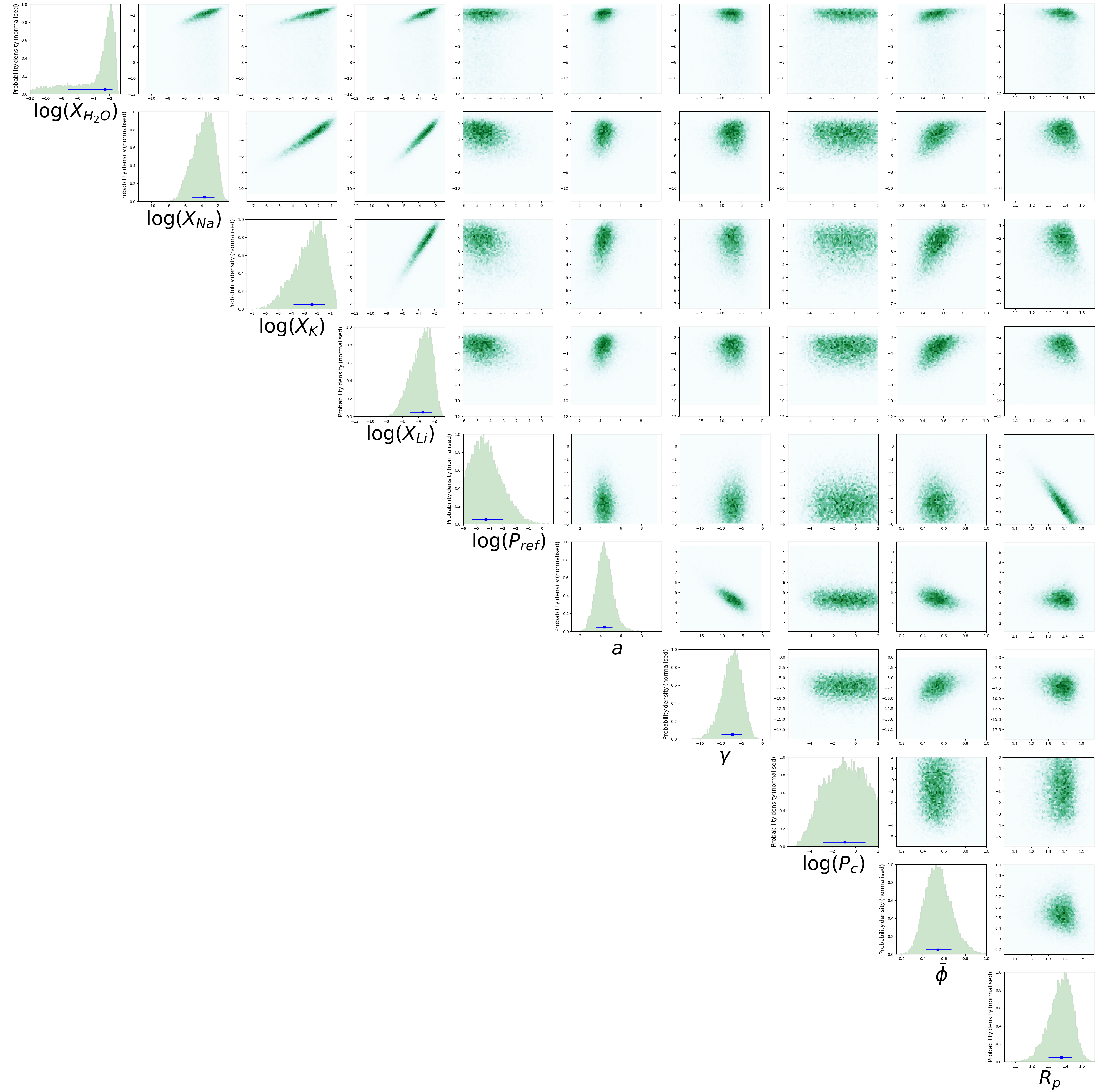}
\caption{Marginalized posterior probability densities for the retrieved species and haze parameters with the reference planet radius as a free parameter. }
\label{fig:posterior_test}
\end{figure*}

The retrieval results presented in Sect.~\ref{sec:retrieval}, Table~\ref{tab:bf_retrieval}, and Fig.~\ref{fig:ptprof} have assumed a reference planet radius of $R_\mathrm{p}=1.37$~$R_\mathrm{Jup}$, and the pressure at this assumed reference radius is a free parameter. We also tested setting the reference planet radius as a free parameter in the retrieval modeling. This additional test resulted in consistent posterior distributions with the original test. The resulting posterior distributions are shown in Fig.~\ref{fig:posterior} and Fig.~\ref{fig:posterior_test}. We present the comparison between these two scenarios in Table~\ref{tab:retrieval_param} along with the prior on each parameter.


\begin{table}[h!]
     \centering
     \caption{Bayesian model comparison detections of atmospheric compositions at the terminator of WASP-127b}
     \label{tab:bf_retrieval}
     \begin{tabular}{lcccc}
     \hline\hline\noalign{\smallskip}
     Model & Evidence  & Best-fit & Bayes factor & Detection \\\noalign{\smallskip}
           & $\ln \mathcal{Z}_i$ & $\chi^2_{r,\mathrm{min}}$ & $\mathcal{B}_{0i}$ & of Ref. \\\noalign{\smallskip}
     \hline\noalign{\smallskip}
     Reference &  619.0 & 1.4 & Ref.   & Ref.    \\\noalign{\smallskip}
     No K &  608.1 & 1.8 & 54838.7 & 5.0$\sigma$  \\\noalign{\smallskip}
     No Na &  612.1 & 1.6 & 1008.3   & 4.1$\sigma$ \\\noalign{\smallskip}     No Li &  614.6 & 1.6 & 79.6   & 3.4$\sigma$ \\\noalign{\smallskip}
     No H$_2$O &  617.8 & 1.5 & 3.4   & 2.1$\sigma$ \\\noalign{\smallskip}
    \hline\noalign{\smallskip}
    \end{tabular}
\end{table}

\begin{table}[h!]
     \centering
     \caption{Prior information and best-fitting retrieval parameters}
     \label{tab:retrieval_param}
     \begin{tabular}{lccc}
     \hline\hline\noalign{\smallskip}
               &       & Retrieval 1 & Retrieval 2 \\\noalign{\smallskip}     Parameter & Prior & Fixed $R_\mathrm{ref}$ & Free $R_\mathrm{ref}$\\\noalign{\smallskip}
               &       & (adopted) & \\\noalign{\smallskip}
     \hline\noalign{\smallskip}
     $\log(X_\mathrm{Na})$       &  $\mathcal{U}(-12,-0.5)$   & $-3.17^{+1.03}_{-1.46}$ &  $-3.49^{+1.20}_{-1.54}$\\\noalign{\smallskip}
     $\log(X_\mathrm{K})$        &  $\mathcal{U}(-12,-0.5)$   & $-2.13^{+0.85}_{-1.32}$ &  $-2.42^{+0.99}_{-1.43}$\\\noalign{\smallskip}
     $\log(X_\mathrm{Li})$       &  $\mathcal{U}(-12,-0.5)$   & $-3.17^{+0.97}_{-1.51}$ &  $-3.48^{+1.15}_{-1.58}$\\\noalign{\smallskip}
     $\log(X_\mathrm{H_2O})$     &  $\mathcal{U}(-12,-0.5)$   & $-2.60^{+0.94}_{-4.56}$ &  $-2.56^{+0.92}_{-4.69}$\\\noalign{\smallskip}
     $\log(P_\mathrm{ref})$ [bar] &  $\mathcal{U}(-6,2)$   & $-4.38^{+0.56}_{-0.42}$ &  $-4.30^{+1.31}_{-1.03}$\\\noalign{\smallskip}
     $\log(P_\mathrm{cloud})$ [bar] &  $\mathcal{U}(-6,2)$   & $-1.02^{+0.20}_{-1.93}$ &  $-0.94^{+1.82}_{-1.93}$\\\noalign{\smallskip}
     $\log(a)$      &  $\mathcal{U}(-4,10)$   & $4.57^{+0.83}_{-0.64}$ &  $4.35^{+0.80}_{-0.77}$\\\noalign{\smallskip}
     $\gamma$       &  $\mathcal{U}(-20,2)$   & $-7.36^{+2.33}_{-2.56}$  & $-7.26^{+2.31}_{-2.60}$\\\noalign{\smallskip}
     $\phi$         &  $\mathcal{U}(0,1)$     & $0.52^{+0.10}_{-0.09}$   & $0.54^{+0.13}_{-0.11}$\\\noalign{\smallskip}
     $R_\mathrm{ref}$ [\Rjup]         &  $\mathcal{U}(0.1,3.0)$  & 1.37 (fixed)   & $1.38^{+0.06}_{-0.08}$\\\noalign{\smallskip}
    \hline\noalign{\smallskip}
    \end{tabular}
\end{table}

\section{Additional figures and tables}\label{app:addtabfig}

\begin{figure*}[h!]
\centering
\includegraphics[width=0.88\linewidth]{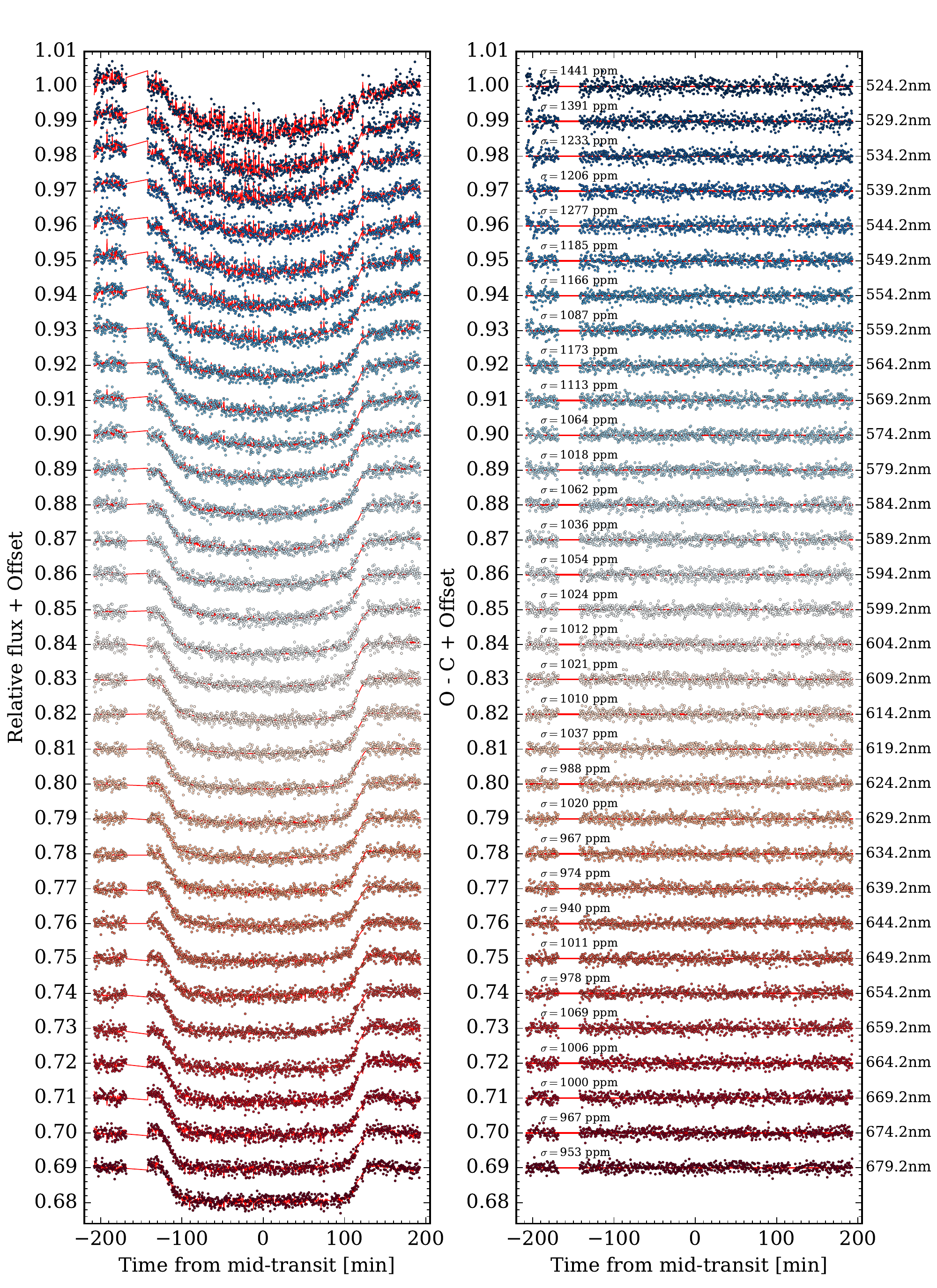}
\caption{Spectroscopic light curves after removing the common-mode systematics ({\it left panel}) and corresponding best-fitting residuals ({\it right panel}) of \object{WASP-127b} obtained with the R1000R grism of GTC/OSIRIS. The passbands have been indicated in Fig.~\ref{fig:GTCSpectra}. This shows the passbands from 524.2~nm to 679.2~nm, which are continued in Fig.~\ref{fig:GTCSpecLC2} and Fig.~\ref{fig:GTCSpecLC3}. \label{fig:GTCSpecLC1}}
\end{figure*}

\begin{figure*}[h!]
\centering
\includegraphics[width=0.88\linewidth]{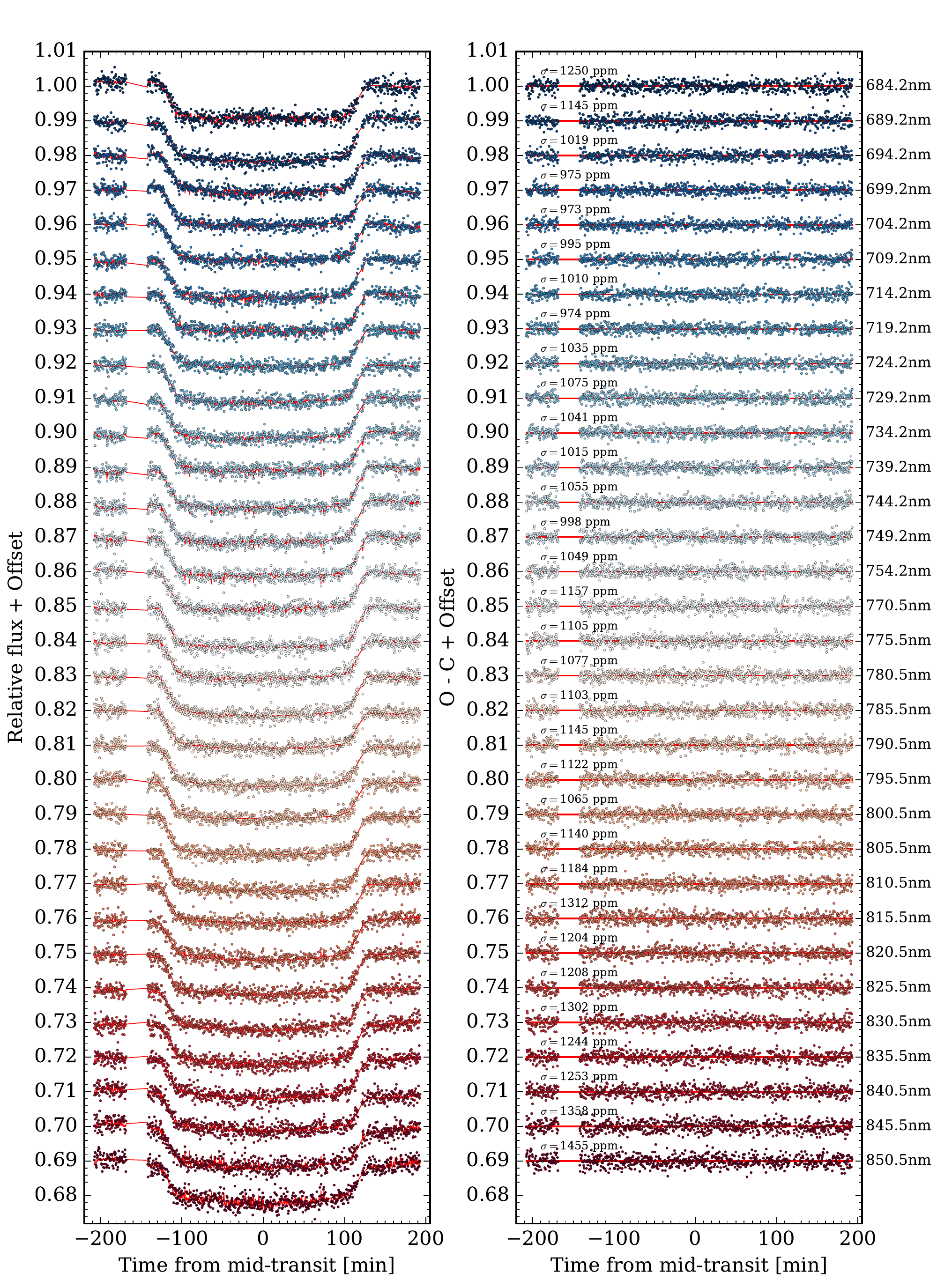}
\caption{Same as Fig.~\ref{fig:GTCSpecLC1}, but for the passbands from 684.2~nm to 850.5~nm. \label{fig:GTCSpecLC2}}
\end{figure*}

\begin{figure*}[h!]
\centering
\includegraphics[width=0.88\linewidth]{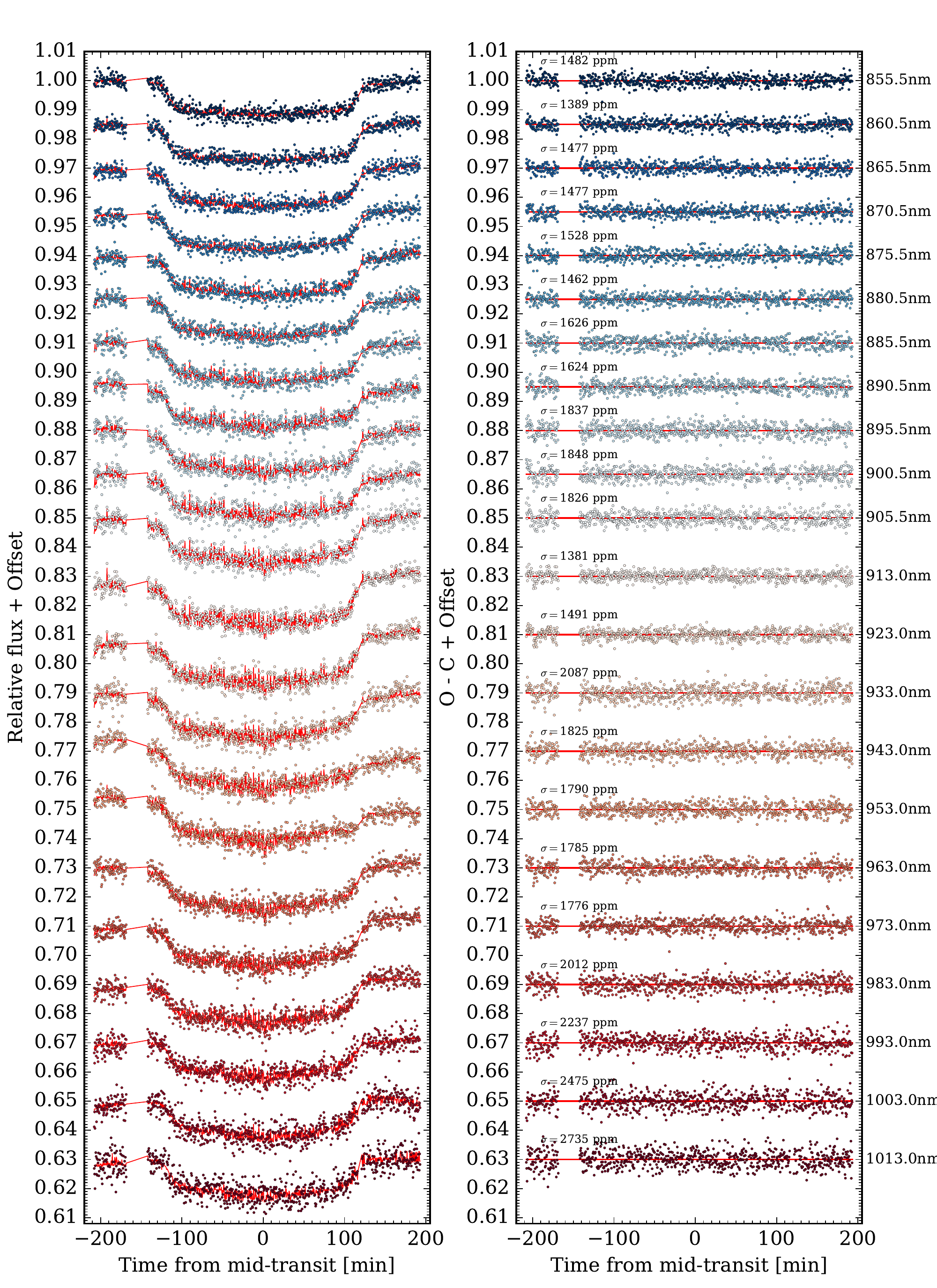}
\caption{Same as Fig.~\ref{fig:GTCSpecLC1}, but for the passbands from 855.5~nm to 1013.0~nm. \label{fig:GTCSpecLC3}}
\end{figure*}

Figures~\ref{fig:GTCSpecLC1}--\ref{fig:GTCSpecLC3} show the GTC/OSIRIS spectroscopic light curves after removing the common-mode systematics and best-fitting light-curve residuals. Table~\ref{tab:not_transpec} shows the transit depths of the NOT/ALFOSC transmission spectrum \citep{2017A&A...602L..15P}, which has been re-analyzed using the method described in this paper. Table~\ref{tab:gtc_transpec} shows the transit depths of the newly derived GTC/OSIRIS transmission spectrum. 

\begin{figure*}
\centering
\includegraphics[width=1\linewidth]{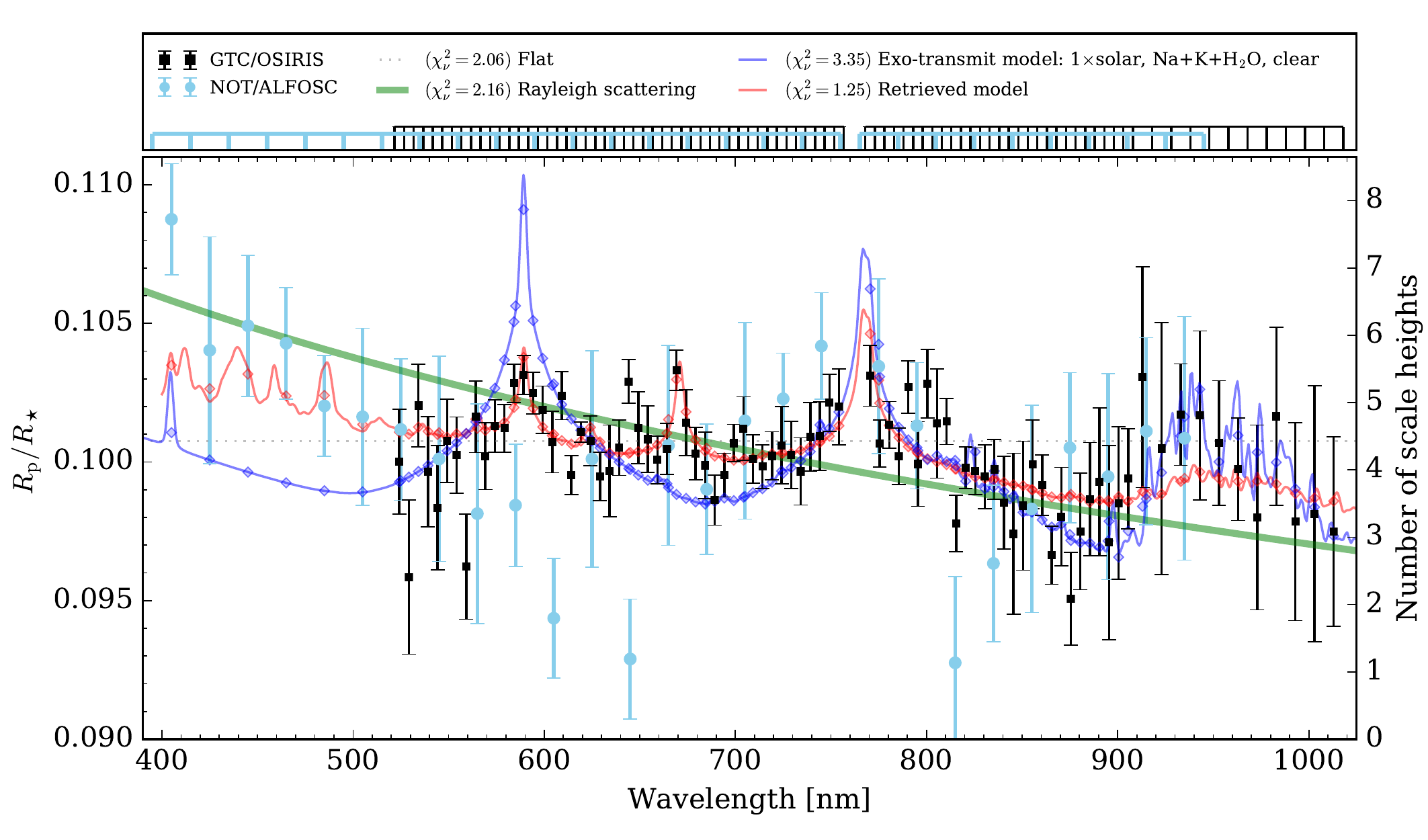}
\caption{Transmission spectrum of \object{WASP-127b}. The GTC/OSIRIS and NOT/ALFOSC measurements are shown in squares and circles with error bars, respectively. A flat line is used to represent the optically thick clouds. The green line shows a representative pure Rayleigh scattering atmosphere. The purple line shows a fiducial atmospheric model with the solar abundance, with Na, K, and H$_2$O included, and with an isothermal temperature of 1400~K \citep{2017PASP..129d4402K}. The red line shows the retrieved model spectrum from our retrieval modeling. The binned models are shown in diamonds. One atmospheric scale height is equivalent to $H_\mathrm{eq}/R_\star=0.00243$. All the reduced chi-squares in this plot only simply consider the offset as a free parameter. \label{fig:GTCTranSpec}}
\end{figure*}

\begin{table}[h!]
     \footnotesize
     \centering
     \caption{Transmission spectrum obtained with NOT/ALFOSC}
     \label{tab:not_transpec}
     \begin{tabular}{cccc}
     \hline\hline\noalign{\smallskip}
     \# & \multicolumn{2}{c}{Wavelength~(\AA)} & \RpRs \\\noalign{\smallskip}
        & Center & Width&  \\\noalign{\smallskip}
     \hline\noalign{\smallskip}
 1 &  4050 &   200 & 0.1128 $\pm$ 0.0018 \\\noalign{\smallskip}
 2 &  4250 &   200 & 0.1081 $\pm$ 0.0040 \\\noalign{\smallskip}
 3 &  4450 &   200 & 0.1093 $\pm$ 0.0020 \\\noalign{\smallskip}
 4 &  4650 &   200 & 0.1084 $\pm$ 0.0020 \\\noalign{\smallskip}
 5 &  4850 &   200 & 0.1061 $\pm$ 0.0018 \\\noalign{\smallskip}
 6 &  5050 &   200 & 0.1055 $\pm$ 0.0029 \\\noalign{\smallskip}
 7 &  5250 &   200 & 0.1049 $\pm$ 0.0019 \\\noalign{\smallskip}
 8 &  5450 &   200 & 0.1041 $\pm$ 0.0038 \\\noalign{\smallskip}
 9 &  5650 &   200 & 0.1021 $\pm$ 0.0041 \\\noalign{\smallskip}
10 &  5850 &   200 & 0.1023 $\pm$ 0.0019 \\\noalign{\smallskip}
11 &  6050 &   200 & 0.0983 $\pm$ 0.0019 \\\noalign{\smallskip}
12 &  6250 &   200 & 0.1043 $\pm$ 0.0039 \\\noalign{\smallskip}
13 &  6450 &   200 & 0.0970 $\pm$ 0.0021 \\\noalign{\smallskip}
14 &  6650 &   200 & 0.1045 $\pm$ 0.0033 \\\noalign{\smallskip}
15 &  6850 &   200 & 0.1032 $\pm$ 0.0023 \\\noalign{\smallskip}
16 &  7050 &   200 & 0.1058 $\pm$ 0.0030 \\\noalign{\smallskip}
17 &  7250 &   200 & 0.1065 $\pm$ 0.0015 \\\noalign{\smallskip}
18 &  7450 &   200 & 0.1083 $\pm$ 0.0019 \\\noalign{\smallskip}
19 &  7750 &   200 & 0.1075 $\pm$ 0.0031 \\\noalign{\smallskip}
20 &  7950 &   200 & 0.1054 $\pm$ 0.0022 \\\noalign{\smallskip}
21 &  8150 &   200 & 0.0967 $\pm$ 0.0031 \\\noalign{\smallskip}
22 &  8350 &   200 & 0.1004 $\pm$ 0.0028 \\\noalign{\smallskip}
23 &  8550 &   200 & 0.1023 $\pm$ 0.0038 \\\noalign{\smallskip}
24 &  8750 &   200 & 0.1046 $\pm$ 0.0027 \\\noalign{\smallskip}
25 &  8950 &   200 & 0.1036 $\pm$ 0.0036 \\\noalign{\smallskip}
26 &  9150 &   200 & 0.1052 $\pm$ 0.0033 \\\noalign{\smallskip}
27 &  9350 &   200 & 0.1051 $\pm$ 0.0044 \\\noalign{\smallskip}
    \hline\noalign{\smallskip}
    \end{tabular}
    \tablefoot{The corrective offset $\Delta\RpRs=0.004056$ has not been subtracted from the values listed in this table.}
\end{table}

\begin{table*}[h!]
     \footnotesize
     \centering
     \caption{Transmission spectrum obtained with GTC/OSIRIS}
     \label{tab:gtc_transpec}
     \begin{tabular}{ccccccccc}
     \hline\hline\noalign{\smallskip}
     \# & \multicolumn{2}{c}{Wavelength~(\AA)} & \RpRs & & \# & \multicolumn{2}{c}{Wavelength~(\AA)} & \RpRs \\\noalign{\smallskip}
        & Center & Width &  & &    & Center & Width &\\\noalign{\smallskip}
     \hline\noalign{\smallskip}
 1 &  5242 &    50 & 0.1000 $\pm$ 0.0019 & & 44 &  7392 &    50 & 0.1009 $\pm$ 0.0012 \\\noalign{\smallskip}
 2 &  5292 &    50 & 0.0958 $\pm$ 0.0028 & & 45 &  7442 &    50 & 0.1009 $\pm$ 0.0012 \\\noalign{\smallskip}
 3 &  5342 &    50 & 0.1020 $\pm$ 0.0015 & & 46 &  7492 &    50 & 0.1021 $\pm$ 0.0010 \\\noalign{\smallskip}
 4 &  5392 &    50 & 0.0996 $\pm$ 0.0020 & & 47 &  7542 &    50 & 0.1020 $\pm$ 0.0014 \\\noalign{\smallskip}
 5 &  5442 &    50 & 0.0983 $\pm$ 0.0022 & & 48 &  7705 &    50 & 0.1031 $\pm$ 0.0011 \\\noalign{\smallskip}
 6 &  5492 &    50 & 0.1008 $\pm$ 0.0015 & & 49 &  7755 &    50 & 0.1007 $\pm$ 0.0008 \\\noalign{\smallskip}
 7 &  5542 &    50 & 0.1002 $\pm$ 0.0014 & & 50 &  7805 &    50 & 0.1013 $\pm$ 0.0008 \\\noalign{\smallskip}
 8 &  5592 &    50 & 0.0962 $\pm$ 0.0019 & & 51 &  7855 &    50 & 0.1002 $\pm$ 0.0010 \\\noalign{\smallskip}
 9 &  5642 &    50 & 0.1016 $\pm$ 0.0013 & & 52 &  7905 &    50 & 0.1027 $\pm$ 0.0009 \\\noalign{\smallskip}
10 &  5692 &    50 & 0.1002 $\pm$ 0.0012 & & 53 &  7955 &    50 & 0.0999 $\pm$ 0.0015 \\\noalign{\smallskip}
11 &  5742 &    50 & 0.1013 $\pm$ 0.0009 & & 54 &  8005 &    50 & 0.1028 $\pm$ 0.0012 \\\noalign{\smallskip}
12 &  5792 &    50 & 0.1012 $\pm$ 0.0009 & & 55 &  8055 &    50 & 0.1014 $\pm$ 0.0020 \\\noalign{\smallskip}
13 &  5842 &    50 & 0.1028 $\pm$ 0.0007 & & 56 &  8105 &    50 & 0.1015 $\pm$ 0.0008 \\\noalign{\smallskip}
14 &  5892 &    50 & 0.1031 $\pm$ 0.0007 & & 57 &  8155 &    50 & 0.0978 $\pm$ 0.0010 \\\noalign{\smallskip}
15 &  5942 &    50 & 0.1025 $\pm$ 0.0008 & & 58 &  8205 &    50 & 0.0998 $\pm$ 0.0008 \\\noalign{\smallskip}
16 &  5992 &    50 & 0.1019 $\pm$ 0.0009 & & 59 &  8255 &    50 & 0.0997 $\pm$ 0.0008 \\\noalign{\smallskip}
17 &  6042 &    50 & 0.1007 $\pm$ 0.0011 & & 60 &  8305 &    50 & 0.0995 $\pm$ 0.0011 \\\noalign{\smallskip}
18 &  6092 &    50 & 0.1024 $\pm$ 0.0009 & & 61 &  8355 &    50 & 0.0997 $\pm$ 0.0011 \\\noalign{\smallskip}
19 &  6142 &    50 & 0.0995 $\pm$ 0.0007 & & 62 &  8405 &    50 & 0.0985 $\pm$ 0.0017 \\\noalign{\smallskip}
20 &  6192 &    50 & 0.1011 $\pm$ 0.0004 & & 63 &  8455 &    50 & 0.0974 $\pm$ 0.0029 \\\noalign{\smallskip}
21 &  6242 &    50 & 0.1008 $\pm$ 0.0009 & & 64 &  8505 &    50 & 0.0984 $\pm$ 0.0023 \\\noalign{\smallskip}
22 &  6292 &    50 & 0.0995 $\pm$ 0.0009 & & 65 &  8555 &    50 & 0.0999 $\pm$ 0.0016 \\\noalign{\smallskip}
23 &  6342 &    50 & 0.0997 $\pm$ 0.0017 & & 66 &  8605 &    50 & 0.0992 $\pm$ 0.0011 \\\noalign{\smallskip}
24 &  6392 &    50 & 0.1005 $\pm$ 0.0010 & & 67 &  8655 &    50 & 0.0966 $\pm$ 0.0010 \\\noalign{\smallskip}
25 &  6442 &    50 & 0.1029 $\pm$ 0.0008 & & 68 &  8705 &    50 & 0.0980 $\pm$ 0.0018 \\\noalign{\smallskip}
26 &  6492 &    50 & 0.1012 $\pm$ 0.0013 & & 69 &  8755 &    50 & 0.0951 $\pm$ 0.0017 \\\noalign{\smallskip}
27 &  6542 &    50 & 0.1008 $\pm$ 0.0012 & & 70 &  8805 &    50 & 0.0975 $\pm$ 0.0021 \\\noalign{\smallskip}
28 &  6592 &    50 & 0.1001 $\pm$ 0.0009 & & 71 &  8855 &    50 & 0.0987 $\pm$ 0.0020 \\\noalign{\smallskip}
29 &  6642 &    50 & 0.1005 $\pm$ 0.0009 & & 72 &  8905 &    50 & 0.0993 $\pm$ 0.0027 \\\noalign{\smallskip}
30 &  6692 &    50 & 0.1033 $\pm$ 0.0007 & & 73 &  8955 &    50 & 0.0971 $\pm$ 0.0035 \\\noalign{\smallskip}
31 &  6742 &    50 & 0.1014 $\pm$ 0.0012 & & 74 &  9005 &    50 & 0.0985 $\pm$ 0.0028 \\\noalign{\smallskip}
32 &  6792 &    50 & 0.1003 $\pm$ 0.0009 & & 75 &  9055 &    50 & 0.0994 $\pm$ 0.0018 \\\noalign{\smallskip}
33 &  6842 &    50 & 0.0999 $\pm$ 0.0012 & & 76 &  9130 &   100 & 0.1031 $\pm$ 0.0040 \\\noalign{\smallskip}
34 &  6892 &    50 & 0.0986 $\pm$ 0.0009 & & 77 &  9230 &   100 & 0.1005 $\pm$ 0.0046 \\\noalign{\smallskip}
35 &  6942 &    50 & 0.0995 $\pm$ 0.0008 & & 78 &  9330 &   100 & 0.1017 $\pm$ 0.0018 \\\noalign{\smallskip}
36 &  6992 &    50 & 0.1007 $\pm$ 0.0007 & & 79 &  9430 &   100 & 0.1017 $\pm$ 0.0030 \\\noalign{\smallskip}
37 &  7042 &    50 & 0.1012 $\pm$ 0.0011 & & 80 &  9530 &   100 & 0.1007 $\pm$ 0.0022 \\\noalign{\smallskip}
38 &  7092 &    50 & 0.1001 $\pm$ 0.0009 & & 81 &  9630 &   100 & 0.0997 $\pm$ 0.0019 \\\noalign{\smallskip}
39 &  7142 &    50 & 0.0998 $\pm$ 0.0008 & & 82 &  9730 &   100 & 0.0980 $\pm$ 0.0033 \\\noalign{\smallskip}
40 &  7192 &    50 & 0.1002 $\pm$ 0.0009 & & 83 &  9830 &   100 & 0.1016 $\pm$ 0.0032 \\\noalign{\smallskip}
41 &  7242 &    50 & 0.1006 $\pm$ 0.0014 & & 84 &  9930 &   100 & 0.0978 $\pm$ 0.0036 \\\noalign{\smallskip}
42 &  7292 &    50 & 0.1002 $\pm$ 0.0012 & & 85 & 10030 &   100 & 0.0981 $\pm$ 0.0046 \\\noalign{\smallskip}
43 &  7342 &    50 & 0.0997 $\pm$ 0.0012 & & 86 & 10130 &   100 & 0.0975 $\pm$ 0.0034 \\\noalign{\smallskip}
    \hline\noalign{\smallskip}
    \end{tabular}
\end{table*}

\end{appendix}

\end{document}